\documentclass[final,5p,times,twocolumn,letterpaper]{elsarticle}

\usepackage{amsthm}         
\usepackage{amsmath} 	   
\usepackage{amssymb}       
\usepackage{amsfonts}
\usepackage{graphicx} 	    
\usepackage{verbatim}
\usepackage{color}
\usepackage{colortbl}
\usepackage{float}
\usepackage{multirow}

\usepackage{subcaption}

\definecolor{babyblue}{rgb}{0.54, 0.81, 0.94}
\definecolor{corn}{rgb}{0.98, 0.93, 0.36}

\begin{document}

\begin{frontmatter}

\title{A new kind of cyclic universe}

\author[add0,add1]{Anna Ijjas} 
\ead{anna.ijjas@cfa.harvard.edu}
\author[add1,add2]{Paul J. Steinhardt
}
\ead{steinh@princeton.edu}

\address[add0]{Institute for Theory and Computation, Harvard-Smithsonian Center for Astrophysics, Harvard University, Cambridge, MA 02138, USA}
\address[add1]{Department of Physics, Princeton University, Princeton, NJ 08544, USA}
\address[add2]{Princeton Center for Theoretical Science, Princeton University, Princeton, NJ 08544, USA}


\date{\today}

\begin{abstract}
Combining  intervals of ekpyrotic (ultra-slow) contraction with a (non-singular) classical bounce naturally  leads to a novel cyclic theory of the universe in which the Hubble parameter, energy density and temperature oscillate periodically, but the scale factor grows by an exponential factor from one cycle to the next.  The resulting cosmology  not only resolves the homogeneity, isotropy, flatness and monopole problems and generates a nearly scale invariant spectrum of density perturbations, but it also addresses a number of age-old cosmological issues that big bang inflationary cosmology does not.  There may also be potential wider-ranging implications for fundamental physics, black holes and quantum measurement.
\end{abstract}


\end{frontmatter}

\section{Introduction}

Historically, dating back to Friedmann and Tolman, cyclic cosmological models have been based on having a Friedmann-Robertson-Walker scale factor  $a(t)$  that oscillates at regularly spaced intervals of time between zero and some large finite value \cite{1922ZPhy...10..377F,1932PhRv...39..835T}. This leads to problems: As $a(t)$ approaches zero, the ordinary matter and radiation densities approach the Planck density and any macroscopic compact objects existing well before the bounce (such as black holes) merge together, potentially obstructing the bounce.  Understanding the currently unknown effects of quantum gravity at these densities becomes critical in determining if and how a (singular) quantum bounce occurs. 

The cyclic model considered here is fundamentally different.   The Hubble parameter $H(t) \equiv \dot{a}/a$, where dot denotes differentiation with respect to time $t$, oscillates periodically between positive and negative values from cycle to cycle, as one might expect.   However, $a(t)$ does not.  Instead, $a(t)$ grows substantially during the usual radiation, matter and dark energy dominated expanding phases, but shrinks very little during the contraction phases.  The result for $a(t)$ is an overall, substantial increase from one cycle to the next.  See Fig.~\ref{fig:1}.  The radically different time-dependence of $H(t)$ and $a(t)$ is a straightforward but inherently general relativistic effect with no simple Newtonian interpretation.

A local observer (like us) judges the evolution to be cyclic because the energy density, temperature, and the concentrations of individual physical quantities (baryon density, dark matter density,  black hole density, etc.) all vary periodically with $H(t)$.  The fact that $a(t)$ has grown by an exponential factor over the previous cycle has no measurable consequences to the local observer who cannot `see' beyond a Hubble radius, $r_H(t) \equiv 1/|H(t)|$, in units where the speed of light is set equal to unity. 
 Globally, the average behavior of $a(t)$ over many cycles is {\it de Sitter-like} with a small effective Hubble parameter,  a factor of ten or so times the current Hubble parameter $H_0$.  The on-average de Sitter-like behavior ensures that entropy and compact objects created in earlier cycles are diluted and, hence, irrelevant in later cycles. In this and other ways explained below, this new cyclic scenario is a curious mix of elements already familiar to cosmologists but rearranged to produce a surprisingly different yet compelling outcome.

First, the new cyclic theory resolves the homogeneity, isotropy, flatness, and monopole problems and generates a nearly scale-invariant spectrum of primordial adiabatic, gaussian density fluctuations without requiring special initial conditions or triggering the kind of quantum runaway that leads to the multiverse effect.  Second, the density perturbations are generated 
without producing a primordial spectrum of tensor fluctuations, a combination that is in agreement with current observations. Third, the evolution of the universe is described to leading order by classical equations of motion at every stage.  Consequently, the theory's outcomes are true predictions in the conventional sense, meaning the theory   
is testable -- making predictions about density fluctuations, cosmic gravitational waves, dark energy, and the stability of the vacuum.   Fourth, the new cyclic theory evades some of the foundational problems of cosmological models based on having a big bang.  The cosmic singularity, cosmic quantum-to-classical transition, and trans-planckian fluctuation problems of earlier theories are avoided.    There may even be intriguing implications for black holes, cosmic censorship, and the quantum measurement problem.

\section{Ingredients}

The necessary ingredients are prosaic: the known forms of matter and radiation, including dark matter; scalar fields, including one responsible for the current cosmic acceleration and the ultra-slow contraction that follows;  three spatial dimensions and ordinary (classical) time.  {\it We emphasize that  we do not invoke extra dimensions, branes, and other elements inspired by string theory for this new approach to cyclic cosmology.}  

Mechanisms for constructing models with a non-singular classical  bounce (see, {\it e.g.}, \cite{Ijjas:2018qbo} and references therein) or ekpyrotic (ultra-slow) contraction \cite{Steinhardt:2001st,Bars:2013vba}, have been discussed extensively elsewhere in other contexts.  
For the purposes of this paper, the microphysics details underlying these mechanisms are not essential.  All one needs to assume is that the mechanisms are possible.  
The novelty is realizing how the elements  combine in a natural evolutionary sequence to produce a scenario with the wide-ranging explanatory and predictive power noted above.

\section{The basic scenario} 

\begin{figure}[tb]
  \centering
\includegraphics[width=.85\linewidth]{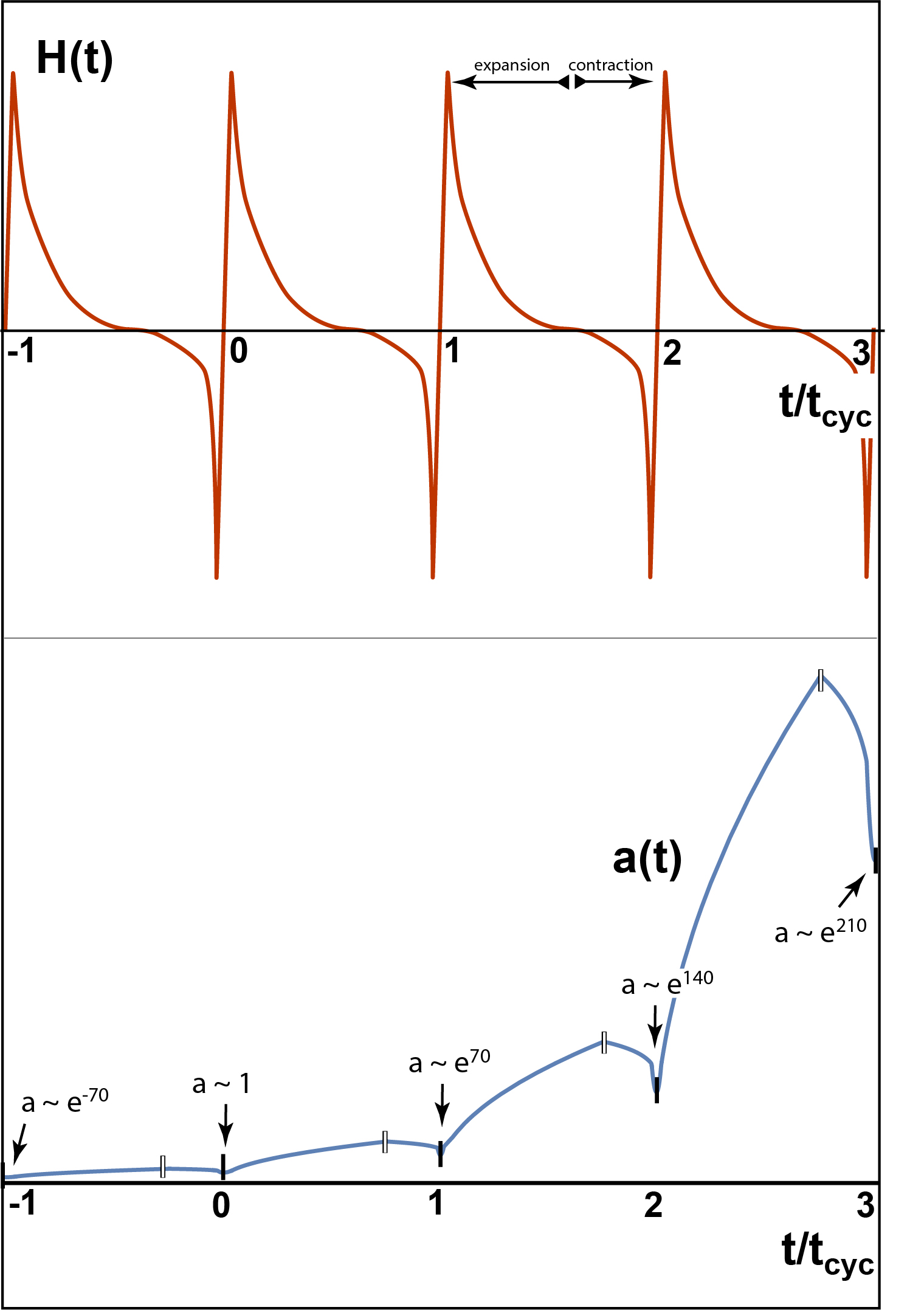}
\caption{In the new cyclic model, the Hubble parameter $H(t)$ oscillates (upper panel) with an amplitude, say, of $\sim 10^{10}$~GeV over a period $t_{cyc}$.  The scale factor $a(t)$, by contrast, is not periodic.  Rather, it goes through long periods of expansion followed by shorter periods of contraction.  The net result is an exponential increase in $a(t)$  over the course of each cycle, producing an on-average de Sitter-like expansion over many cycles.  
The thin filled rectangular
 tickmarks indicate the cross-overs from contraction to expansion due to a (non-singular) classical bounce. The thin unfilled rectangular tickmarks indicate the cross-overs from accelerated expansion to ekpyrotic (ultra-slow) contraction . 
 See also Fig.~\ref{fig:2}.
}
\label{fig:1}
\end{figure}

Arranging for $H(t)$ to oscillate between large positive and large negative values while $a(t)$ increases from cycle to cycle, as shown in Fig.~\ref{fig:1}, is not a  contrivance.  It is an automatic consequence of general relativity in a cyclic model that combines ekpyrotic  (ultra-slow) contraction and (non-singular) classical bounces. According to the Friedmann equations, $a(t) \sim |t|^{1/\varepsilon}$  and $|H(t)|$ is proportional to $a^{-\varepsilon}$ during a phase with equation-of-state $\varepsilon$.  By definition, the equation-of-state is $\varepsilon_\pm \equiv (3/2) (1+ p/\varrho)$, where $p$ is the pressure and   $\varrho$ is the energy density.   We will henceforth add the subscript $-$ when referring to the value of $\varepsilon$ during the contracting ($H<0$) phase and  a subscript $+$ to indicate the value  during the expanding ($H>0$) phase. 

During the expanding phase of a cycle, $H(t)$ begins large and positive and decreases.  The universe undergoes periods in which the  dominant form of energy density is radiation ($\varepsilon_+=2$), matter ($\varepsilon_+=3/2$), and dark energy ($\varepsilon_+ \approx 0$), just as in conventional big bang cosmology.  Because $\varepsilon_+ = {\cal O}(1)$ and the expansion phase lasts a long time,  it is possible for  $a(t) \sim |t|^{1/\varepsilon_+}$ to increase and for $H(t) \sim a^{-\varepsilon_+}$ to decrease by large exponential factors by the time the expanding phase ends.  

The expanding phase ends and the ekpyrotic contracting phase begins when $H(t)$ passes below zero and continues to decrease, as described in Sec.~\ref{sec:3.1} below.  
An ekpyrotic  phase corresponds to   $\varepsilon_- \gg 1$.   Because $\varepsilon_- \gg 1$, the magnitude of $H(t)$, which is  proportional to $a^{-\varepsilon_-}$, can increase by an exponential factor while $a(t) \sim |t|^{1/\varepsilon_-}$ may only shrink by a factor of ${\cal O}(1)$.   

During a (non-singular) classical bounce phase,  $H(t)$ rapidly increases from a large negative value to a large positive value of roughly the same magnitude while $a(t)$ is roughly unchanged.  At that moment, the universe enters the next expansion phase and a new cycle begins.
We note that, recently, the first example of a non-singular classical bounce has been worked out that is locally well-behaved to perturbative order, overcoming problems of earlier models; for details see \cite{Ijjas:2017pei}.

The net result over the course of a full cycle is that: ($i$) $H(t)$ oscillates between exponentially large positive and negative values; and, ($ii$)  $a(t)$ grows exponentially during the expansion phase but decreases very little during the contraction and bounce phases, resulting in an overall exponential increase in $a(t)$ by the end of the cycle, as shown in Fig.~\ref{fig:1}.

\subsection{A single cycle of evolution: an illustrative example}
\label{sec:3.1}

The bounce ensures that the evolution of the universe is dominantly classical at all stages of the cycle.  In particular, as $H(t)$ oscillates, it never reaches close to the Planck mass, $\sim 10^{19}$~GeV.   Likewise, the Hubble radius, $r_H(t) \equiv 1/|H(t)|$, never reaches the Planck length, $\sim 10^{-33}$~cm, and the energy density associated with the scalar field driving the ekpyrotic (ultra-slow) contraction does not reach close to the Planck density $\sim (10^{19} \,{\rm GeV})^4$.  As an illustrative example that satisfies all known quantitative constraints, we choose the minimum value of $r_H$ to be $\sim 10^{-25}$~cm, corresponding to a maximum $|H| \sim 10^{10}$~GeV and a maximum energy density of $\varrho \sim (10^{15} \,{\rm GeV})^4$.  

If $r_H$ is $\sim 10^{-25}$~cm and the temperature is $T \sim 10^{15}$~GeV at the start of the expansion phase, the scale factor $a(t)$ increases by a factor $f \sim e^{120/\varepsilon_+} \sim e^{60}$ by the time the temperature reaches today's cosmic microwave background temperature, where here we have taken account of the duration of the radiation, matter and dark energy dominated phases.     Over this same period, the Hubble radius increases by  a factor of $a^{\varepsilon_+} = f^2 \sim e^{120}$ from a microscopic size of $10^{-25}$~cm to the current value of the Hubble radius $10^{28}$~cm.

That takes us up to the present epoch.  The current dark energy dominated phase may continue for a period into the future, but, in a cyclic universe, it must eventually terminate.  For example.
the dark energy may be due to a quintessence-like scalar field $\phi$  with scalar field potential $V(\phi)$. During the radiation- and matter-dominated phases, $\phi$ is frozen at a value where $V(\phi)>0$ by Hubble friction.  The field begins to roll downhill only when the dark energy comes to dominate, which is only recently at $H(t) \sim H_0$, the present value of the Hubble parameter. The accelerated expansion phase ends and contraction begins when $\phi$ rolls from $V(\phi)>0$ to $V(\phi)<0$, which may be a few Hubble times.    At this point, the scalar field changes from a quintessence-like field that drives accelerated expansion to an ekpyrotic field that governs the subsequent phase of ultra-slow contraction.

{\it Note that the dark energy density plays several roles in the new cyclic theory}:  it sets the time when dark energy first comes to dominate, which is approximately the current Hubble time $H_0$;  it sets the characteristic time scale (up to a modest numerical factor) for the duration of the expanding phase; it also sets the contracting phase that follows (see below), and, hence, the total period of a cycle.
For our example, let's suppose that the current accelerated expansion lasts an additional 10 Hubble times into the future or, equivalently, 10 $e$-folds of increase in $a(t)$ before acceleration ends and  the ekpyrotic (ultra-slow) contraction phase begins.  In this case, the total cycle lasts $O(10) \, H_0^{-1}$.  
   
The equation-of-state for a  homogeneous scalar field with canonical kinetic energy density is given by
 \begin{equation}
 \varepsilon_-  =
3\times \frac{ \frac{1}{2} \dot{\phi}^2}{\frac{1}{2} \dot{\phi}^2+V(\phi)} \,. 
 \end{equation}
For the case of a negative exponential potential $V = -  V_0 e^{ \phi/M}$ (where $V_0>0$ is constant and we use reduced Planck units $ 8 \pi G_{\rm N}=1$ for Newton's gravitational constant $G_N$), the equation-of-state during the contracting phase can be shown to be nearly constant with $\varepsilon_- = 1/(2 M^2)$.   It is not difficult to reach large values of $\varepsilon_-$; for example, for $M=0.1$, the value  is $\varepsilon_- = 50$.  
     
The ekpyrotic contraction phase with $\varepsilon_- \gg 1$ plays two important roles in the cyclic scenario.  First, it is a remarkably powerful smoothing and flattening mechanism, the only mechanism currently known that not only smooths classically but also when quantum fluctuations are included.
 
Classically,  because the ekpyrotic scalar field energy density is proportional to $1/a^{2 \varepsilon_-}$ and all other components (matter, radiation, dark energy, gradient energy, spatial curvature, anisotropy) scale as $1/a^{2q}$ where $q\ll  \varepsilon_-$, the homogeneous ekpyrotic field energy grows overwhelmingly faster than all the other components as $a(t)$ contracts, driving the universe towards an ultra-uniform, ultra-flat, ultra-local state exponentially dominated by scalar field energy as the bounce approaches.  The larger the value of $\varepsilon_-$, the faster and more powerful is the classical smoothing and flattening effect. 
 The extraordinarily rapid convergence to homogeneous and isotropic  Friedmann-Robertson-Walker conditions has been demonstrated  beginning from wildly nonlinear initial conditions using numerical general relativity \cite{Garfinkle:2008ei}.

Quantum mechanically, quantum fluctuations  in the ekpyrotic field produce {\it decaying mode} adiabatic curvature fluctuations \cite{Creminelli:2004jg} that do not interfere with the classical smoothing.   This is to be contrasted with  inflaton fields that generically produce {\it growing mode} scalar and tensor perturbations.  Growing modes are the root cause of  B-modes with substantial amplitudes (which have not been observed), quantum runaway, and the multiverse effect.  Because the ekpyrotic contraction phase produces decaying modes, it does not have these problematic features.  (See Sec.~\ref{sec:4} for a discussion of how a nearly-scale-invariant spectrum of density perturbations is generated.)

The second critical role of an ekpyrotic contraction phase with $\varepsilon_- \gg 1$ is that it enables
  the magnitude of $H(t)$ (which proportional to   $ a^{\varepsilon_-}$) to grow by a large factor of $f^2 \sim e^{120}$ to its original value  during a period in which $a(t)$ shrinks  by a factor of $e^{120/\epsilon_-} = {\cal O}(10)$. 
Because  $a(t)$ hardly decreases during the contraction phase,  the density of matter and radiation hardly increases  during the contraction phase. Likewise, the physical distance between black holes and other compact objects hardly decreases at all.   There is no issue of compact objects created in earlier cycles merging and disrupting the contraction and bounce phases. This aspect is unlike cyclic models of the past.  The only significantly growing component is the scalar field energy density, which exponentially dominates all other forms energy density by the time the ekpyrotic contraction phase ends. 


Finally, the cycle completes with a (non-singular) classical bounce.
During this phase, $H(t)$ rapidly increases from $-f^2 \sim e^{120}$ to $+f^2 \sim e^{120}$, returning to the value it had at the beginning of the cycle (just after the previous bounce).   The large energy density stored in the ekpyrotic scalar field is converted to matter and radiation through a reheating process analogous to that assumed in inflation (see, {\it e.g.}, \cite{Kofman:1994rk}).  The value of $H(t)$ and the density of matter and radiation have returned to the values they had a cycle ago and 
 a new period of oscillation in $H(t)$ begins.  Over this same cycle, $a(t)$ has grown by an exponential factor of $  e^{10} f  \sim e^{70} \sim 10^{30}$ during the expanding phase and shrunk by only a factor of  $e^{120/\epsilon_+} = {\cal O}(10)$ during the contraction phase.

\subsection{Multiple cycles and de Sitter-like behavior}

\begin{figure}[tb]
  \centering
\includegraphics[width=.85\linewidth]{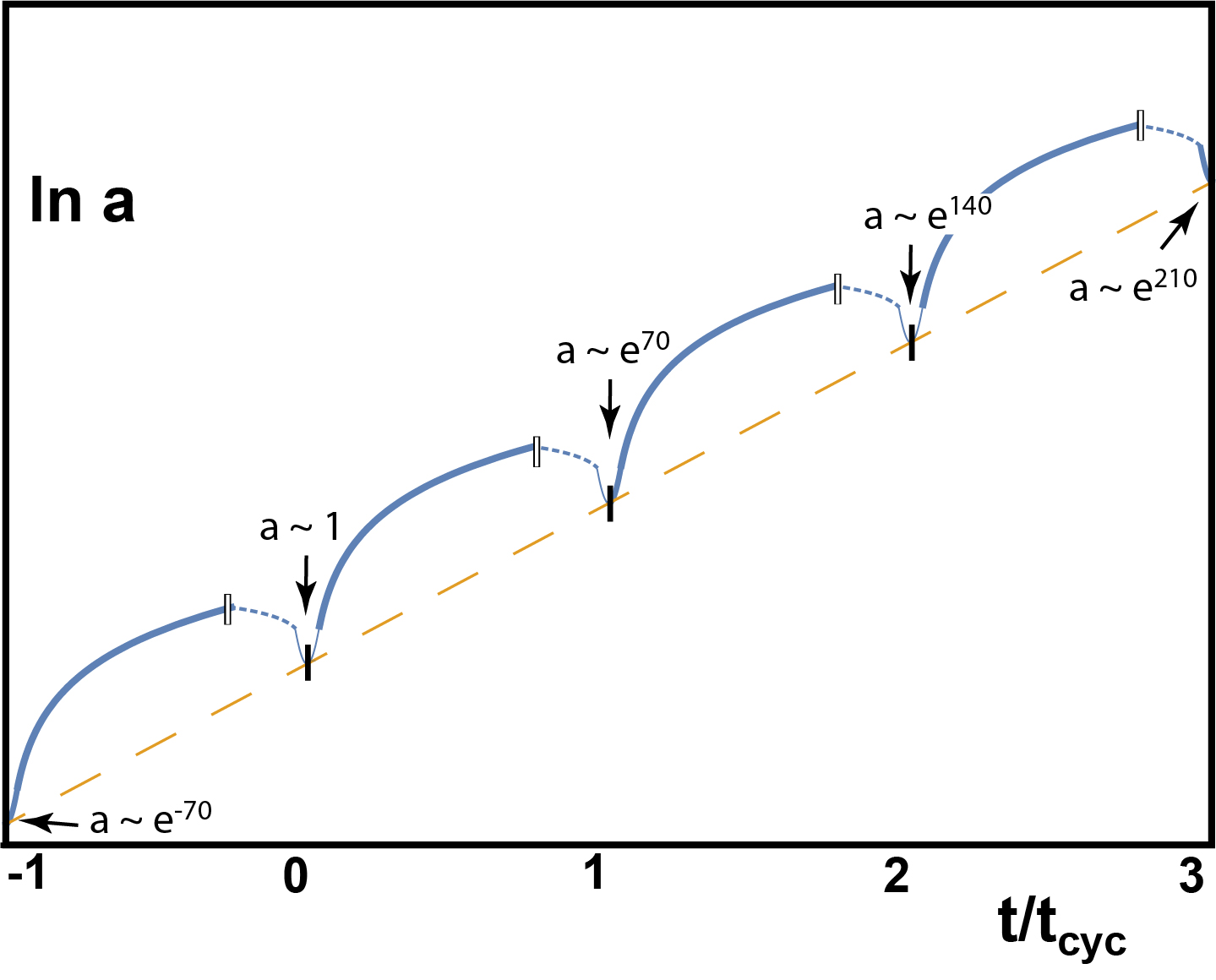}
\caption{
A plot of $\ln a(t)$ vs. $t$ showing the behavior of the scale factor and the three phases that repeat during each cycle: the bounce phase ($\sim 10^{-35}$~sec, thin solid line), the expanding phase ($\sim 10^{11}$~yr, thick solid line), and the contracting phase ($\sim  10^{9}$~yr, dashed line), where the numbers of parentheses correspond to the illustrative example discussed in the text. The result is a net exponential increase in $a(t)$ from one cycle to the next, resulting in an on-average exponential (de Sitter-like) expansion rate over many cycles (dashed line). 
 In our example, the dashed line corresponds to $a(t) \sim \exp \bar{H}(t)$ where $\bar{H}$ is ${\cal O}(10)H_0$ and $H_0$ is the current value of the Hubble parameter.
}
\label{fig:2}
\end{figure}

 Fig.~\ref{fig:2} illustrates the three basic phases (bounce, expansion and contraction) that repeat from cycle to cycle.
To estimate the times for each of these phases, recall that, in our example, the Hubble radius $r_H $   grows by a factor of $f^2$ during the time interval $\Delta t$ between the beginning of the expansion phase and today.  According the Friedmann equation, $r_H= \varepsilon_+ \Delta t$.   Then, during the contraction phase, $r_H$  shrinks by a factor of $f^2$ during an interval of time $\Delta_t' = (\varepsilon_+/\varepsilon_-) \Delta t$.  The difference in $\varepsilon$ in the expansion and contraction phases means that the duration of the contracting phase is shorter than the duration of the expansion phase.  Hence, if the expansion phase lasts 150 billion years, as in our example (assuming an additional 10 $e$-folds of dark energy expansion in our future before contraction begins), then the ensuing ekpyrotic contraction phase might last, say, less than 10 billion years.  This modest difference is indicated in Figs.~\ref{fig:1} and~\ref{fig:2}.

Because of the overall exponential increase in $a(t)$ over an entire cycle, the volume of space that grows to the size of the current observable universe ($\sim H_0^{-1}$) a cycle from now is microscopic today, about $ 10^{-25}f$~cm or about 10~cm, a bit bigger than the diameter of a baseball.  

{\it That is,  an empty volume of space equal to the diameter of a baseball today will evolve to the size of our observable universe ($\sim H_0^{-1}\sim 10^{28}$~cm) a cycle from now.}  This can be understood as follows. Today, the size of the observable universe is about the current Hubble radius ($\sim H_0^{-1}$), exponentially larger than a baseball.    
During contraction, the baseball-sized volume  ($\propto a(t)$) will hardly change, but, 
due to the ekpyrotic scalar field, 
the Hubble radius ($\propto a^{\varepsilon_-}$)  will shrink by a factor of $f^2$ to about $ 10^{-25}$~cm.  At this point, the Hubble radius has become exponentially smaller than a baseball, by a factor of $f$.  During the expansion phase, ordinary radiation and matter domination causes  the Hubble radius to grow again by $a^{\varepsilon_+} \sim f^2$; during the same period, the baseball-sized volume of space has grown by $f$.   The two are the same. This is what we mean when we say a baseball-sized volume today will evolve to the size of the current Hubble radius one cycle from now. 

The repeated exponential increase in $a(t)$ cycle after cycle leads to a surprising effect.  Namely, the scale factor follows an on-average de Sitter-like exponential growth curve, as shown in Fig.~\ref{fig:2}.  This means 
that the energy, matter, and entropy created in earlier cycles is  exponentially diluted from one cycle to the next such that only an infinitesimal fraction is  observable within a Hubble radius during the subsequent expanding phases.     

The  matter and radiation observed today was created during or just after the most recent bounce, {\it i.e.,} the event that occurred about 14 billion years ago and that has often been attributed to the big bang.  More specifically, as we have explained above, during each period of ekpyrotic contraction, the gravitational blueshift effect naturally pumps exponentially large amounts of energy into the ekpyrotic scalar field that drives the dark energy and ultra-slow contraction phases.  After the ekpyrotic contraction phase completes, a reheating process converts scalar field energy density to ordinary matter and radiation, producing the same energy densities after the bounce as in previous cycles. Anything we observe today is created from the new matter and radiation, effectively resetting the horizon to equal the Hubble radius at reheating.  

{\it The second law is obeyed:} the total entropy of the universe increases; but the entropy observed within the Hubble radius is the same from cycle to cycle.  Entropy, black holes, matter and radiation  from earlier cycles lie beyond our horizon, spread thinly over the vast expanses of space created as a result of the  overall exponential increases in $a(t)$ that occur from cycle to cycle.

\section{Cosmological consequences}
\label{sec:4}

The cyclic theory described in the previous section combines features of earlier cosmological models.
The behavior of $H(t)$ is cyclic.  Within each cycle is a conventional hot big bang-like expansion phase.  A smooth exponentially increasing curve can be drawn through the minima of $a(t)$, which would describe an expanding de Sitter phase with constant $H = {\cal O}(10 H_0)$; see Fig.~\ref{fig:2}.  But note that  this on-average de Sitter expansion occurs at an exponentially lower value of $H$ compared to inflation and so plays no role in smoothing or flattening the universe or in generating density perturbations.  Rather, it is the period of ekpyrotic (ultra-slow) contraction that is responsible for these features.  

Although many of the building-block ideas are familiar, the way they are put together produces a scenario that resolves cosmological problems while avoiding the pitfalls of earlier approaches and shedding new light on some long-standing puzzles.
The essential feature is that {\it the evolution of the universe through  all stages  is dominantly classical}.  Quantum corrections are always small.  That means there is no period of quantum domination at any time during cosmic evolution: {\it no big bang, no quantum-determined initial conditions,  no quantum runaway that leads to the multiverse effect}.  
All coarse-grain properties of the universe are deterministically set by the governing classical equations.  Fine details, such as the precise distribution of small density perturbations after the bounce, are determined by random quantum fluctuations, but their statistical properties are set by classically-determined coarse-grain properties of the universe. 

This condition is made possible by having   
a  (non-singular) classical bounce, which requires a classical  violation of the {\it null convergence condition},  
$R_{\alpha \beta} n^{\alpha} n^{\beta} \ge 0$,
for all null vectors $n^{\alpha}$, where $R_{\alpha \beta}$ is the Ricci tensor.  This can be achieved through an appropriate modification of Einstein gravity at high energy densities near the bounce ($\varrho \sim (10^{15} \,{\rm GeV})^4$) or stress-energy that violates the null energy condition or both.   A well-developed set of examples is based on a modification of Einstein gravity described in Horndeski and Galileon theories that results in braiding the scalar field and extrinsic curvature \cite{Ijjas:2017pei,Ijjas:2016vtq}.  The modification introduces, among other things, an amendment to the Friedmann equation that is proportional to a combination of $H$ and $\dot{\phi}$.  

During most of the cycle, $H$ and $\dot{\phi}$ are so small that the braiding effect is negligible because the two 
quantities are both individually small. They are only significant during the contracting phase, as   $H$ becomes increasingly  negative and $\dot{\phi}$ increases due to gravitational blue shift.  When the braiding term becomes non-negligible, the bounce phase begins. After the bounce, the universe expands, $H$ and $\dot{\phi}$ decrease, and the braiding term becomes insignificant again.  

It is appropriate to view the classical bounce as a solution to the {\it cosmic singularity problem}.  Just as an event horizon shields the outside observer from the time-like singularity within a black hole, the bounce shields the universe from a space-like cosmic singularity.   In this sense, the bounce can be considered as an extension of cosmic censorship to cosmological singularities.

The {\it causal horizon problem} is immediately resolved by replacing the big bang with a bounce preceded by a period of contraction.  The {\it homogeneity, isotropy and flatness problems} are resolved if the preceding period consists of a long enough period of ekpyrotic (ultra-slow) contraction where `long enough' means that the $H(t)$ shrinks by at least 60 $e$-folds.  Ekpyrotic contraction means that $H(t)$ shrinks a lot while $a(t)$ shrinks comparatively less; in this case, we emphasize that there can be hardly any shrinkage in $a(t)$ at all.  If the reheating temperature is sufficiently low ($<10^{16}$~GeV, say), the {\it monopole problem} is resolved because the monopoles would be too massive to be abundantly created.  

The {\it generation of density fluctuations} occurs during the ekpyrotic contraction phase. The scale factor $a(t)$ is nearly constant but $r_H = |H^{-1}|$ is shrinking so fast that  $a /r_H$ grows exponentially. In this way, quantum fluctuations generated on sub-horizon scales ($k > a/r_H$)  become super-horizon modes ($k \ll a/r_H$) as contraction proceeds. Notably,  the fluctuations and the Hubble radius are much larger than the  Planck scale at all stages.  Unlike inflation, there is no issue of trans-planckian fluctuations. 
    
As noted above, an ekpyrotic phase with a single scalar field generates adiabatic curvature fluctuations that decay in amplitude during the contracting phase; for the same reason,  primordial tensor (gravitational wave) fluctuations are not generated \cite{Boyle:2003km}.  However,  
the  {\it nearly scale-invariant spectrum of curvature  fluctuations} observed in the cosmic microwave background can be straightforwardly generated if there is a second scalar field through the well-known {\it isocurvature} mechanism that has been described elsewhere \cite{Lehners:2007ac,Buchbinder:2007ad}.  Namely, the second field generates a scale-invariant spectrum of  isocurvature (a.k.a. entropic) fluctuations that are converted into curvature fluctuations after the ekpyrotic phase is completed.

\begin{figure}[tb]
  \centering
\includegraphics[width=.8\linewidth]{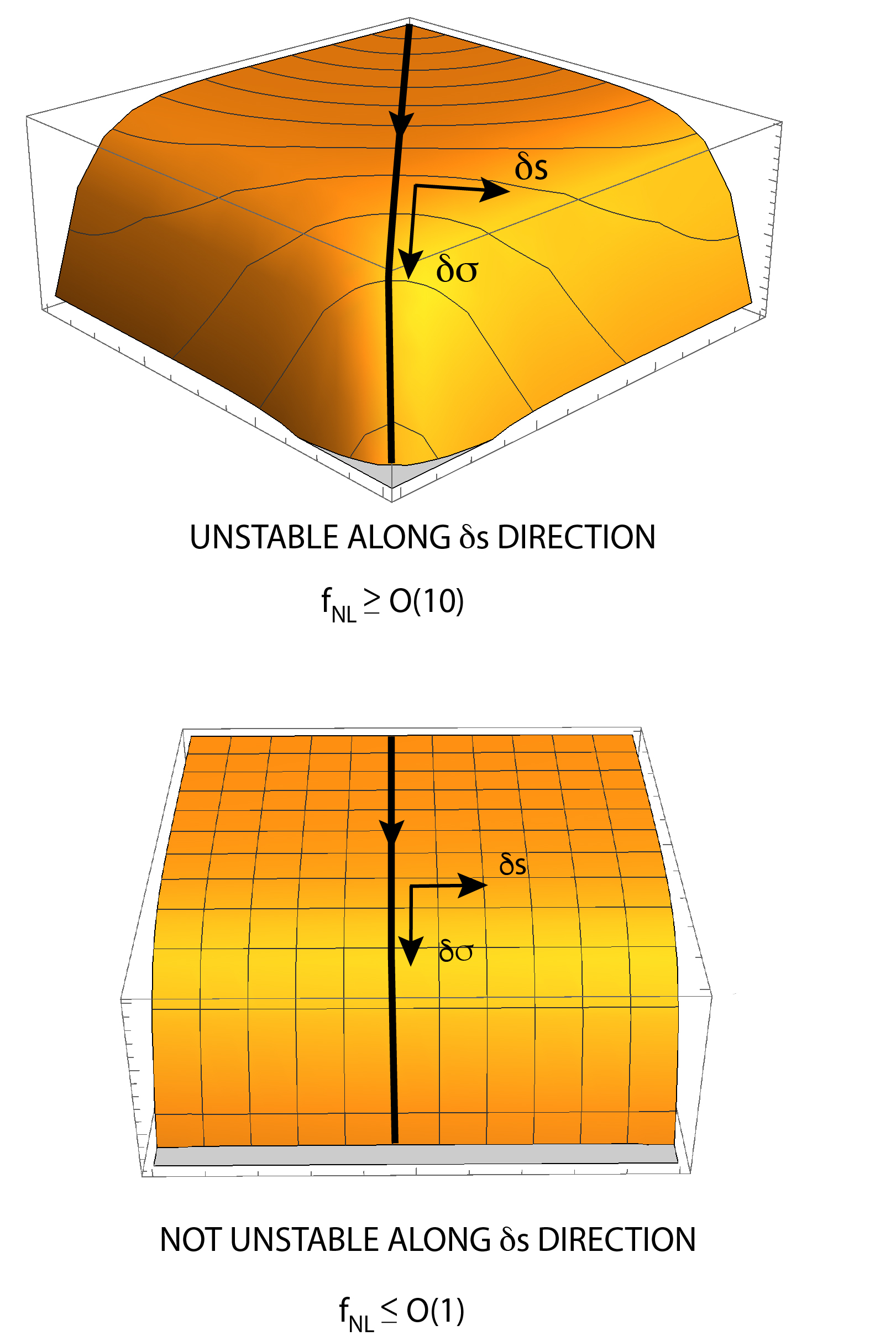}
\caption{The generation of perturbations during an ekpyrotic phase in a model with two scalar fields.  One linear combination ($\sigma$)  sets the trajectory along the potential (solid black line) which, in turn, sets the background equation-of-state.  Quantum fluctuations along the trajectory $\delta \sigma$ decay during a contracting phase and are therefore irrelevant.  Another linear combination of fields ($s$) experiences quantum fluctuations $\delta s$ that generate a nearly scale-invariant spectrum of entropic fluctuations that are converted to curvature perturbations after the ekpyrotic phase completes.  In some examples, the field trajectory is unstable along the $s$-direction (upper panel), in which case the curvature fluctuation spectrum can have significant non-gaussianity ($f_{NL}\ge {\cal O}(10)$).  However, the instability makes these models unsuitable for repeated cycling.  Ekpyrotic phases with no instability (lower panel) are compatible with repeated cycling; these are cases in which  the non-gaussianity is negligible ($f_{NL}\le {\cal O}(1)$). 
}
\label{fig:3}
\end{figure}
There is a subtlety regarding {\it gaussianity} that is worth noting since it has caused some confusion in the literature.   In some (in fact, historically, the originally proposed) models of ekpyrosis \cite{Tolley:2007nq}, the predicted density fluctuations have a significantly non-gaussian distribution.  This occurs, for example, if the ekpyrotic phase is dominated by two scalar fields, both with canonical kinetic energy and both with exponentially steep potentials.
In terms of the standard (local) non-gaussianity parameter $f_{NL}$, the prediction is $f_{NL} \ge {\cal O}(10)$, which is marginally inconsistent with current observations \cite{Lehners:2013cka}.

However, non-gaussianity is {\it not} generic.   It only occurs in cases like the example cited, where the ekpyrotic field trajectory is inherently unstable to small quantum fluctuations away from the ideal  path, as illustrated in the upper panel of Fig.~\ref{fig:3}.  A model like this is poorly suited for cyclic theories because the initial conditions at the beginning of each ekpyrotic phase must be repeatedly finely-tuned in order for the ekpyrotic phase  during each cycle to last long enough.

Simple ekpyrotic models with completely stable field trajectories, as illustrated in the lower panel of Fig.~\ref{fig:3} \cite{Levy:2015awa}, avoid the instability.  In these models, only one field has an exponentially steep potential and the other has a non-linear sigma model-type kinetic coupling to the first.  The kinetic coupling strongly damps the classical motion of the second field as the first rolls downs its potential.  The result is a stable trajectory.   Models like these with stable trajectories are the natural choice for scenarios with repeated cycling because they do not require fine-tuning of initial conditions at the beginning of each ekpyrotic phase.  The stability has an important observational consequence for cyclic models:  {\it the fluctuation spectrum  is predicted to be very nearly gaussian, $f_{NL} \le {\cal O}(1)$}, in accordance with current observations. 

As for {\it the B-mode problem}, we have explained that primordial tensor fluctuations are not generated during ekpyrotic contraction for the same reason that adiabatic density fluctuations are not. Furthermore, the isocurvature mechanism does not generate primordial tensor fluctuations. This feature is in agreement with current observations that have not (yet) detected primordial $B$-modes.  However, there is not simply a null prediction.  Secondary gravitational waves created when density fluctuations re-enter the horizon are predicted to be the leading-order contributions to the tensor fluctuation spectrum with tensor-to-scalar ratios $r \lesssim 10^{-6}$ \cite{Baumann:2007zm}. While current experiments are not sensitive to detect these modes, they may be detectable in the future.

Yet another generic prediction is that the current vacuum and accelerated expansion are temporary.   That is, the vacuum is either unstable and slowly changing with time; or metastable, protected by an energy barrier that will eventually be bypassed through quantum tunneling.  The possibility that we live in an unstable or metastable phase is well-motivated.  In fact, in many formulations of unified theories, including supersymmetric and string theories, there are strong arguments to suggest that the global minimum has negative vacuum energy density (see, {\it e.g.}, \cite{Agrawal:2018own}).  This is just what is called for.  If there exist vacua with negative energy density, classical or quantum mechanics will eventually find it.  The current accelerated expansion phase  
cannot be eternal.  The universe must eventually transit to the true vacuum and begin to contract. Will that bring the universe to a crunch (an end) or a bounce? We have shown a bounce back to an unstable or metastable vacuum with positive energy density is possible \cite{Ijjas:2016vtq,Ijjas:2017pei}.  And if it could happen once, it would necessarily repeat because there would remain the same state of negative energy density and the same laws of classical and quantum mechanics returning the universe to it.   That is, {\it if we conclude from future observations that there was a bounce in our past, there are good reasons to believe there will be another in our future.}

Stepping back, we would argue that an appealing aspect of the cyclic scenario is that it relies on a {\it natural progression} of equations-of-state that make it straightforward to fit together the different stages of evolution.  By `natural progression,' we mean that, as the universe expands, the evolution sequence should go from stages dominated by  high pressure (large $\varepsilon_+$) components  to stages dominated by  low pressure (smaller $\varepsilon_+$)  components because the energy density decreases as $1/a^{2 \varepsilon}$.   For example, in the cyclic picture, expansion begins at about the time when  the ekpyrotic scalar field decays predominantly to radiation ($\varepsilon=2$), which gives way to matter ($\varepsilon=3/2$), and finally to dark energy ($\varepsilon\approx0$).  

During the contracting phase, the reverse should hold.  The natural progression is from low pressure to high.  The dark energy phase ($\varepsilon\approx 0$) ends and contraction begins  when the scalar field causing it rolls or decays to negative potential energy.  The value of $\varepsilon$ grows as the universe contracts to the bounce.  The progression is natural.   One might contrast this with the case of a big bang universe that begins hot (moderate  positive pressure), settles into an  inflationary phase (negative pressure), which then reverts to radiation (moderate positive pressure);  the initial conditions problem in inflation is related in part to this non-monotonic pressure sequence \cite{Gibbons:2006pa}.

Table I summarizes the generic features described here.
\begin{table}
\begin{center}
\begin{tabular}{|c|}
\hline
GENERIC FEATURES OF THE NEW CYCLIC THEORY \\
\hline 
resolves the causal horizon problem \\
resolves the homogeneity and isotropy problems \\
resolves the flatness problem  \\
resolves the monopole problem \\
generates super-horizon-scale fluctuations \\
no transplanckian problem $\spadesuit$ \\
classical (coarse-grain) deterministic evolution at all stages $\spadesuit$ $\dagger$ \\
no cosmic singularity problem $\spadesuit$ $\dagger$\\
no big bang $\spadesuit$ \\
no cosmic quantum-to-classical transition $\spadesuit$ $\dagger$ \\
no chaotic mixmaster problem $\spadesuit$ \\
no initial conditions problem $\spadesuit$ \\
natural progression of equation-of-state $\spadesuit$\\
no quantum runaway (a.k.a. no multiverse)$\spadesuit$\\
no detectable {\it primary} tensor fluctuations $\spadesuit$ \\
predicts  {\it secondary} tensor modes ($r \lesssim 10^{-7}$) $\spadesuit$ \\
predicts current vacuum is unstable or metastable $\spadesuit$\\
 \hline 
\end{tabular}
\caption{$\spadesuit$ Marks features of the cyclic model that are {\it not} claimed features of  big bang inflationary cosmology; $\dagger$ marks features that are {\it not} shared by ekpyrotic cyclic models with quantum (singular) bounces.}
\end{center}
\end{table}
The list  should give one pause.  The first five items are features that inflation is often credited as possessing based on its original formulation.   It was not appreciated until later that obtaining sufficient inflation depends sensitively on initial conditions or that quantum fluctuations lead to quantum runaway and the multiverse effect.  When these features are taken into consideration, it is not clear what inflation resolves or predicts since the opposite outcome is equally possible according to current understanding. 

By contrast, precisely because evolution in the cyclic scenario is dominantly classical throughout, all of the items  in Table I are undeniable features or testable predictions.   This sets a new standard for cosmological models and provides a strong argument for non-singular classical bounces.

\section{Further thoughts}

These results lead us to some admittedly highly speculative remarks inspired  -- but not required -- by  the cyclic scenario presented here.  

The idea that it is possible to shield the universe from reaching a stage where quantum physics dominates over classical may be more than a feature of a particular cosmological model.  One could imagine that it is a generic fundamental principle of nature, analogous to the cosmic censorship conjecture for black holes.  We might call this a {\it generalized cosmic censorship} principle and suggest that it must be satisfied by the ultimate theory that unifies gravity with the other fundamental forces.  

The idea that Einstein's theory of general relativity has to be modified at very high energies is generally accepted, but the conjecture suggests specific properties that the modifications must satisfy.  Non-singular classical bounces that shield the universe from reaching the Planck density would be only one example.  There should be analogous bounce-like behavior  approaching the center of a black hole before reaching the point where quantum gravity effects would be expected to dominate in Einstein gravity.  This has implications for the last stages of black hole evaporation, the information paradox and the possibility of gravitationally stable wormholes.  

 Speculating further, the notion that cosmological evolution is dominantly classical at all stages has implications for the quantum measurement problem and gravity's possible role in resolving it: It suggests that there exists a combination of scales -- curvature $\sim r_H$, mass $\sim r_H^{-1}$ and energy density $\sim H^2$ -- below which a physical system (the universe, compact objects, etc.) can be characterized by a set of measurable quantities (observables) that always have definite values and can always be determined with no (or negligible) effect on the state or dynamical evolution.
This combination of scales could be understood as a gravitational lower bound for macro-realism, {\it i.e.}, the claim that `a physical system which can obtain several macroscopically
distinct states exists in exactly one of its possible states at any time' \cite{Leggett:1985zz}.

Whether one finds these speculations appealing or not, they demonstrate how identifying the correct story of cosmic evolution can impact fundamental physics generally.  The advantage of cosmology is that there exists the technology today to test theories, whereas visiting the interior of a black hole, say, is not advisable.  Showing that it is possible in principle to combine ekpyrotic contraction and classical bounces to obtain a cosmological model with the properties listed in Table I is a significant step forward, but here we have only focused on qualitative features.  At this point,  the quantitative properties are obtained by fixing parameters, as is the case for other cosmological theories.  The next challenge is to further develop the cyclic scenario in conjunction with fundamental physics to go beyond showing that the quantitative conditions are possible and explain why they are likely to occur.   

{\it Acknowledgements.} 
We thank J. Miralda-Escude, J. Ostriker. L. Page, J. Peebles and J. Simons for useful comments.
A.I. is supported by the Simons Foundation `Origins of the Universe Initiative' grant number 550202.  The work of P.J.S. is supported by the DOE grant number DEFG02-91ER40671 and by the Simons Foundation grant number 548512.

\bibliographystyle{apsrev}
\bibliography{new_cyc}

\end{document}